\documentclass{PoS}

\newcommand{\Neff}{\ensuremath{N_{\rm{eff}}}}
\newcommand{\DNeff}{\ensuremath{\Delta\Neff}}
\newcommand{\sinem}{\ensuremath{\sin^2 2\vartheta_{e\mu}}}
\newcommand{\dmsbl}{\ensuremath{\Delta m^2_{41}}}

\title{Light sterile neutrinos and pseudoscalar interactions in Cosmology}

\ShortTitle{Light sterile neutrinos and pseudoscalar interactions in Cosmology}

\author{\speaker{Stefano Gariazzo}\\
       University and INFN, Torino\\
       E-mail: \email{gariazzo@to.infn.it}}

\abstract{
The Short BaseLine (SBL) neutrino oscillations anomalies hint
at the presence of a sterile neutrino with a mass of around 1 eV.
However, such neutrino is highly incompatible with the cosmological data,
in particular from the Cosmic Microwave Background (CMB),
if no new physics is assumed.
An interesting possibility for reconciling the 1 eV sterile neutrino presence
in cosmology is related to the existence of a new pseudoscalar interaction.
If the sterile neutrinos experience such a pseudoscalar interaction,
the cosmological analyses of the full CMB data prefer a sterile neutrino mass
that is fully compatible with the SBL determinations.
The additional interaction allows to obtain also an improved compatibility
of the cosmological predictions with the local measurements
of the Hubble parameter.}

\FullConference{Neutrino Oscillation Workshop\\
		4 - 11 September, 2016\\
		Otranto (Lecce, Italy)}

\begin{document}

\section{Light Sterile Neutrinos}

The so-called short-baseline (SBL) neutrino oscillation anomalies
(see e.g.\ \cite{Gariazzo:2015rra})
can be explained with the introduction of a third square mass difference \dmsbl,
that implies the existence of at least four neutrino mass eigenstates.
Since the number of active flavor neutrinos that experience interactions
with the standard model particles is three, the new eigenstate must correspond
to a new sterile neutrino.
Being \dmsbl\ close to 1~eV$^2$,
this new neutrino is called
``light sterile neutrino''~%
\footnote{We use the approximation
$m_s\simeq m_4\simeq\sqrt{\dmsbl}$,
since the mixing angles $\vartheta_{i4}$ are small
and $m_i\ll m_4$ (for $i=1,2,3$).},
and its properties can be studied using
several cosmological observables, including the Cosmic Microwave Background (CMB)
radiation \cite{Ade:2015xua}.

CMB data are robust enough to obtain precise constraints on the energy density
of relativistic particles in the early universe,
parameterized through the effective number \Neff.
Recent calculations including the full analytic collisional terms confirmed
that the contribution of the standard active neutrinos
to \Neff\ is very close to 3, being it slightly larger because
the decoupling of the neutrinos is not instantaneous:
$\Neff=3.046$ \cite{Mangano:2005cc, deSalas:2016ztq}.
The light sterile neutrino also contributes to this number,
and the calculations show that its contribution $\DNeff=\Neff-3.046$
should be very close to 1, if one considers $m_s\simeq1$~eV
and the best-fit mixing angles obtained from the SBL analyses
\cite{Hannestad:2012ky, Mirizzi:2012we}.
In the standard parameterization, hence, we would have $\Neff\simeq4$, if
the light sterile neutrino exists.

\begin{figure}
\begin{center}
  \includegraphics[width=0.85\linewidth]{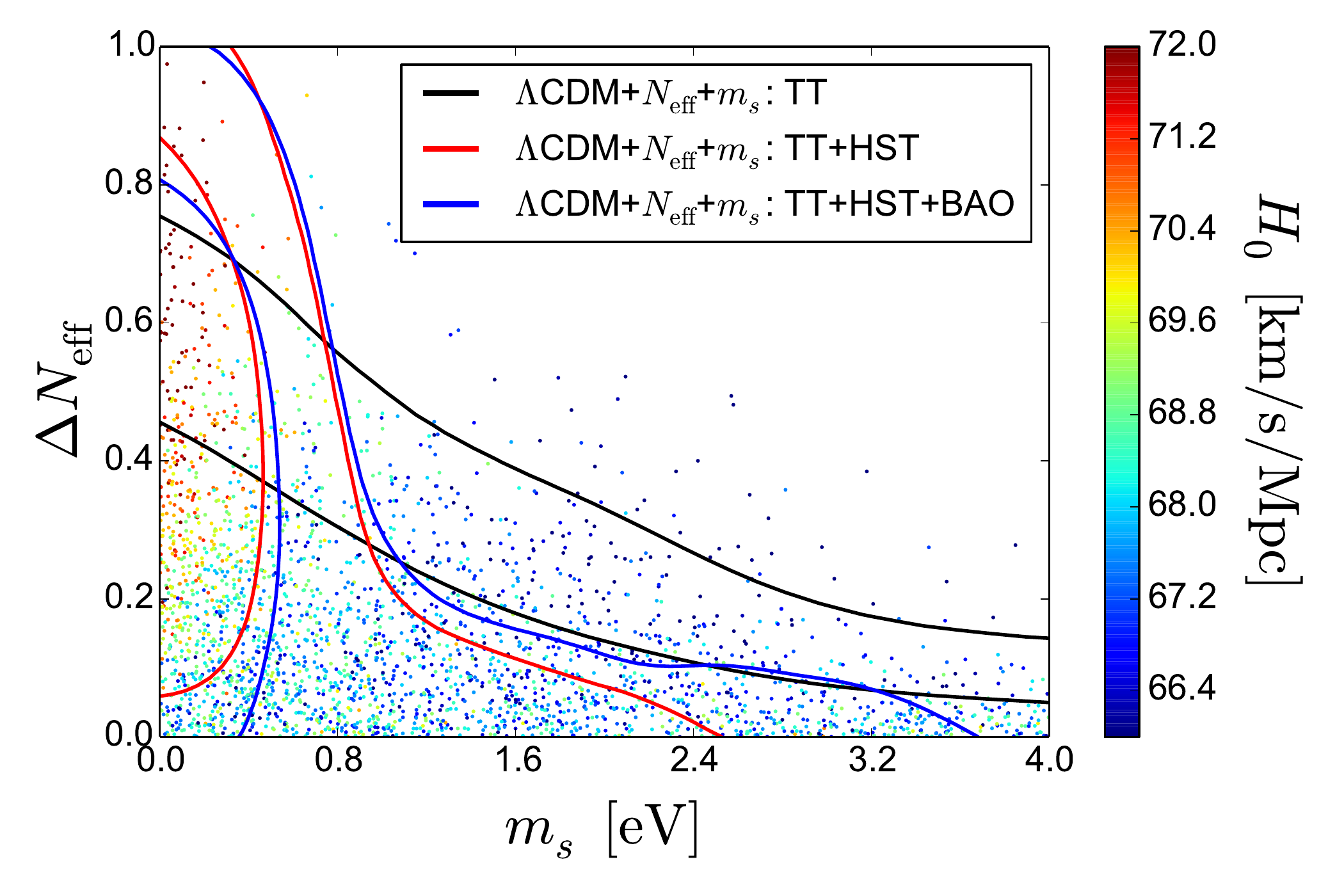}
\end{center}
\label{lsn_nnu_mnus}
\caption{
The cosmological constraints on the light sterile neutrino mass $m_s$ and
the contribution to the effective number $\DNeff$,
obtained with different cosmological data combinations.
The points are color coded by the corresponding value
of the Hubble parameter $H_0$.
From Ref.~\cite{Archidiacono:2016kkh}.
}
\end{figure}

The most recent constraints on \Neff\ 
show that there are no deviations from the standard value 3.046
\cite{Ade:2015xua, Archidiacono:2016kkh, Peimbert:2016bdg},
especially if one considers an additional neutrino with mass around 1~eV
(see Fig.~\ref{lsn_nnu_mnus},
where different cosmological data sets have been considered).
If a light sterile neutrino exists,
its presence in the early universe
must be suppressed by some new mechanism.

\section{Pseudoscalar interaction}

One interesting possibility is to assume that a new interaction exists
in the sterile sector.
The role of the new interaction is to suppress the oscillations in the early
universe and to prevent the thermalization of the light sterile neutrino.
Here we consider a new interaction mediated by a new pseudoscalar boson
\cite{Archidiacono:2014nda, Archidiacono:2016kkh}
that is nearly massless.
If the coupling between the sterile neutrino and the pseudoscalar
is strong enough, the oscillations in the early universe are blocked
by the matter effect driven by the pseudoscalar fluid and
\Neff\ can be smaller than 4.

\begin{figure}
\begin{center}
  \includegraphics[width=0.65\linewidth]{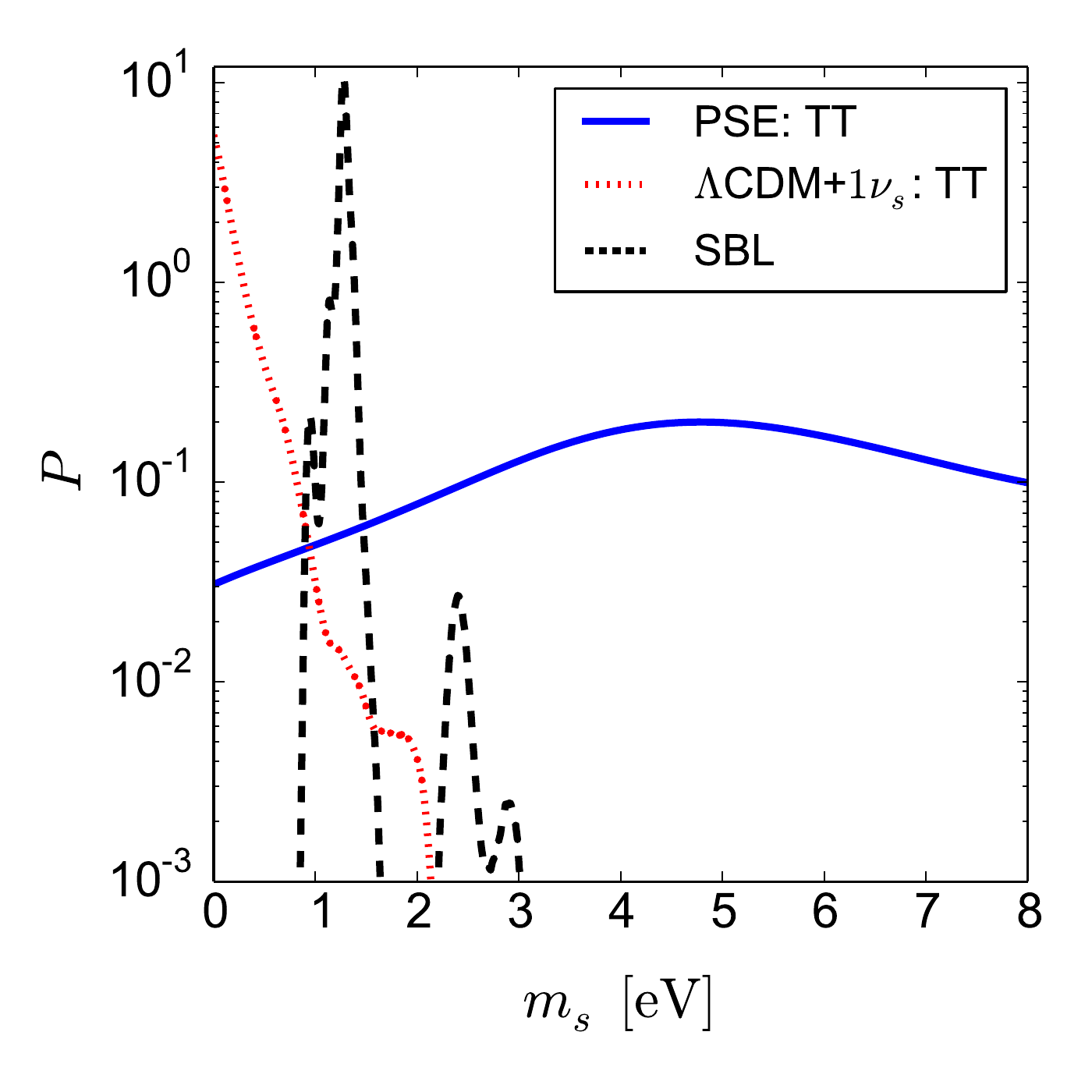}
\end{center}
\label{pseudoR}
\caption{%
\textit{Right panel}: the posterior distribution of
the light sterile neutrino mass $m_s$ as obtained
from the SBL analyses (black dashed) is compared with 
the results obtained from cosmological analyses
that use the standard light sterile neutrino
parameterization with $\DNeff=1$ (red dotted)
or the pseudoscalar model (blue solid).
From Ref.~\cite{Archidiacono:2016kkh}.
}
\end{figure}

At late times, the sterile neutrino decays and populates the pseudoscalar fluid.
In this way, the cosmological bounds on the sterile neutrino mass do not apply.
Indeed, in our analyses we find that a large value for \Neff\ is allowed also
for a sterile neutrino with a mass larger than 1~eV.
As we can see in Fig.~\ref{pseudoR},
the preferred value for $m_s$ is around 5~eV.

\begin{figure}
\begin{center}
  \includegraphics[width=0.85\linewidth]{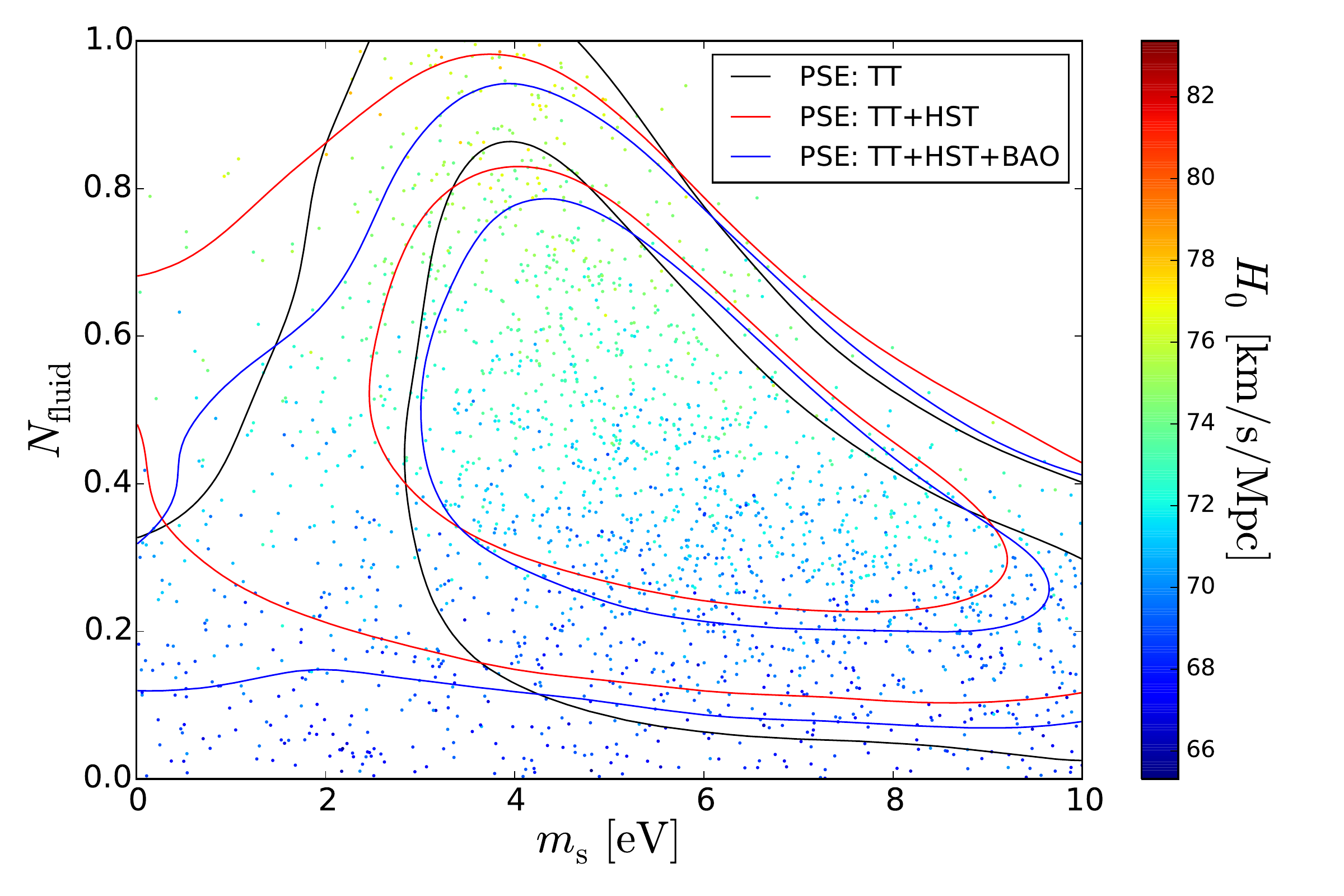}
\end{center}
\label{pseudoL}
\caption{
the same as in Fig.~1,
but within the pseudoscalar model.
From Ref.~\cite{Archidiacono:2016kkh}.
}
\end{figure}

Comparing Fig.~\ref{lsn_nnu_mnus} and \ref{pseudoL},
we can also see how the presence of
the sterile neutrino -- pseudoscalar interaction
allows to have larger values for the Hubble parameter $H_0$,
reconciling the cosmological estimations
with the local measurements \cite{Riess:2016jrr}.

\section{Joint analyses}
The Bayesian analysis mechanism allows to perform combined analyses
of the cosmological and SBL data.
In the past, usually the SBL constraints on $m_s$ were considered
as a prior in the cosmological analysis.
In Ref.~\cite{Archidiacono:2016kkh}, for the first time,
the cosmological posterior probability distribution on $m_s$ has been used
as a prior in the SBL analysis.

\begin{figure}
\begin{center}
  \includegraphics[width=0.7\linewidth]{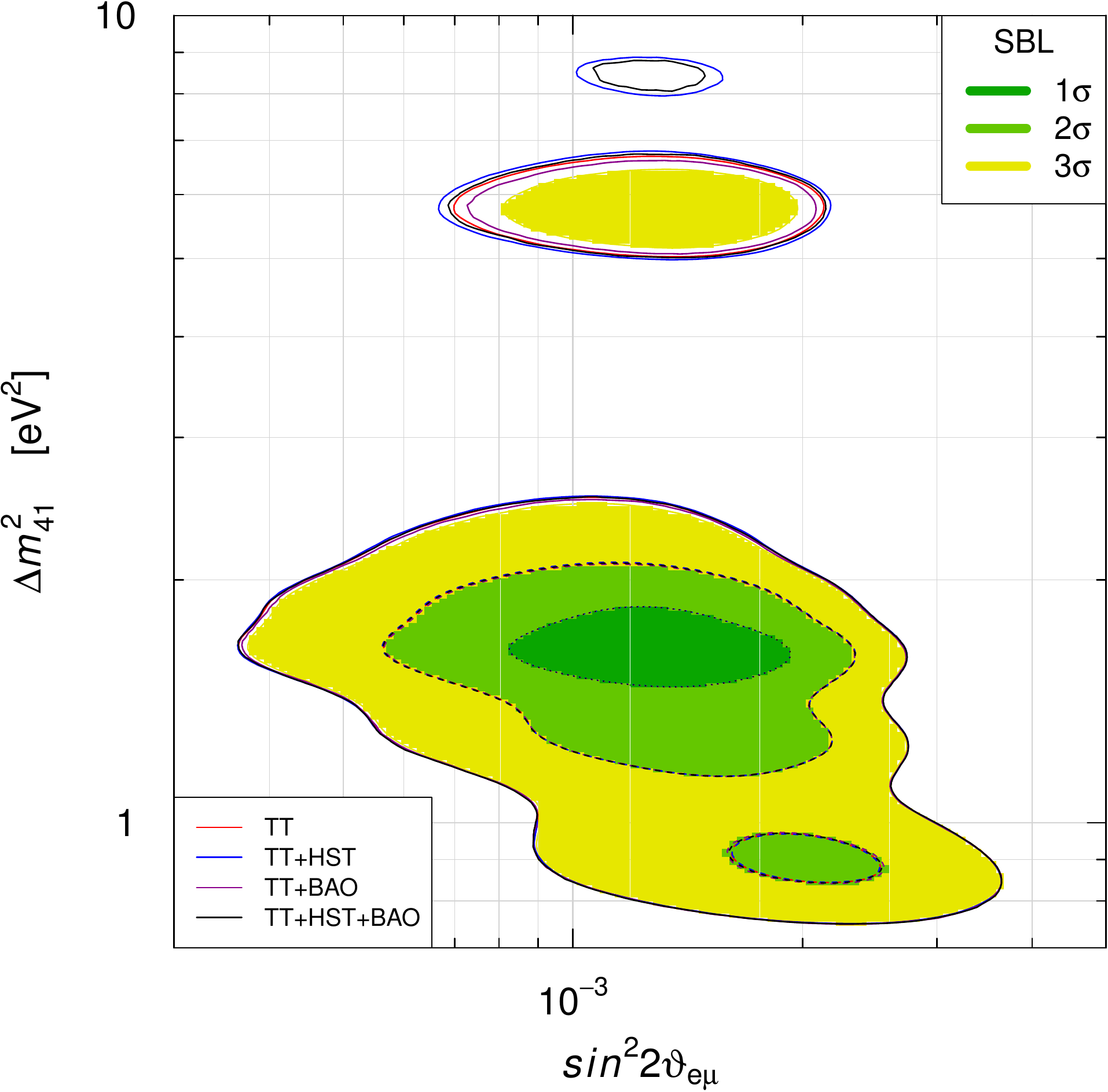}
\end{center}
\label{joint}
\caption{%
Results of the SBL analysis alone (filled regions)
compared with the joint SBL + cosmological data analyses (colored contours),
in the (\sinem, \dmsbl) plane.
From Ref.~\cite{Archidiacono:2016kkh}.
}
\end{figure}

The joint results in the (\sinem, \dmsbl) plane are shown in Fig.~\ref{joint}.
We can see that the SBL data are much stronger in constraining
the sterile neutrino mass, and that the addition of the cosmological information
only has an impact on the 3$\sigma$ regions.
When the cosmological data are included, indeed,
these 3$\sigma$ regions are modified by 
an enlargement of the $\dmsbl\simeq6\,\rm{eV}^2$ regions
and by the appearance of new allowed regions at $\dmsbl\simeq8.5\,\rm{eV}^2$.
These shifts agree with the recent results of IceCube and MINOS,
which prefer a sterile neutrino heavier than 1~eV.

\providecommand{\href}[2]{#2}\begingroup\raggedright\endgroup

% \bibliography{all}
% \bibliographystyle{JHEP}

\end{document}